\documentclass[aps,prd]{revtex4}
\input{epsf}

\begin{document}

\title{The asymptotic quasinormal mode spectrum of non-rotating black holes}
\author{N. Andersson and C.J. Howls}

\address{Department of Mathematics, University of Southampton\\ 
Southampton, SO17 1BJ, UK}

\begin{abstract}
A conjectured connection to quantum gravity 
has led to a renewed interest in highly damped black hole 
quasinormal modes (QNMs). In this paper we present simple derivations
(based on the WKB approximation) of conditions that determine
the asymptotic  QNMs for both Schwarzschild and 
Reissner-Nordstr\"om black holes. This confirms recent results obtained by 
 Motl and Neitzke, but our analysis fills several gaps left by their 
discussion. We study the Reissner-Nordstr\"om results in some detail, 
and show that, in contrast to the asymptotic QNMs of a Schwarzschild
black hole, the Reissner-Nordstr\"om QNMs are typically not
periodic in the imaginary part of the frequency. This leads to the charged 
black hole having peculiar properties which complicate an interpretation
of the results.      
\end{abstract}

\maketitle

\section{Background and motivation}

\subsection{The quasinormal modes}

Black holes oscillate. The associated 
quasinormal modes (QNMs) of oscillation 
are relevant for many reasons. Most importantly, 
numerical relativity has provided ample evidence that 
the QNMs dominate the gravitational-wave signal associated with 
many processes involving dynamical black holes (such as the formation 
of black holes in gravitational collapse or binary merger). 
Since the QNMs encode information 
concerning the parameters of the black hole one may hope that the 
gravitational-wave detectors that are now coming into operation 
will be able to 
use these signals to investigate the black hole population of the 
Universe. 

Even though most studies of QNMs have been motivated by their potential
astrophysical relevance, there are several other reasons why one might 
be interested in understanding the spectrum of oscillations
of a black hole. In particular, the modes have played a key role in discussions 
of black hole stability \cite{whiting}.
A closely related issue concerns mode-completeness. It now seems 
clear that the QNMs do not form a complete set (at least not in the 
conventional sense) because of the presence of power-law
tails caused by the backscattering of waves (see \cite{naprd} 
for a discussion and references). 
The problem also has interesting computational aspects. 
While the slowly damped QNMs, for which $|\mbox{Re }\omega M| >>
|\mbox{Im }\omega M|$, are relatively easy to compute,
highly damped modes present a challenge. 
The main difficulty concerns the fact that, in the frequency domain, 
the QNM eigenfunctions (which represent purely outgoing waves at 
spatial infinity and purely ingoing waves crossing the event horizon) 
grow exponentially. This means that one must, in principle, achieve 
exponential precision in order to impose the boundary conditions. 
The first reliable calculation of high QNM overtones was performed
by Leaver in the mid-1980s using  continued fractions \cite{leaver1}. 
An alternative approach to the problem proceeds via analytic 
continuation using complex coordinates, borrowing standard asymptotic
techniques from quantum mechanics. In particular, 
this was the fundamental idea behind the numerical phase-amplitude
method that has been used to calculate high precision QNMs of both 
Schwarzschild and Reissner-Nordstr\"om black holes \cite{pam1,pam2}. 
Both these results, and those of Nollert and Schmidt \cite{ns}, 
agreed with those of Leaver.

While the reliability of Leaver's method is now well established,
it was not without controversy a decade ago. The discussion concerned
the  behaviour of the QNMs in the limit of very high 
damping. Leaver's results (for the first fifty modes of a Schwarzschild 
black hole) indicated that the modes would asymptotically 
behave as 
$$
\omega_n M \sim 0.08 - { i \over 4} (2n+1) \quad \mbox{ as } n\to \infty
$$ 
This was contradicted by 
results obtained by Guinn et al using a WKB formula \cite{guinn}. Their 
calculation suggested that the real part of the QNM 
frequencies would vanish asymptotically. The controversy was resolved by 
two calculations which agreed that the correct asymptotic result
was 
\begin{equation}
\omega M \sim 0.04371235 - { i \over 4} (2n+1) \quad \mbox{ as } n\to \infty
\end{equation}
The first calculation was based on a 
slight reformulation of the continued fraction algorithm \cite{noll}, 
while the second used a high order phase-integral formula \cite{al,nacqg}. 
The latter calculation also shed light on the reasons for the breakdown 
of the method used by Guinn et al. \cite{guinn}. This issue is 
further discussed in Refs.~\cite{pimrev,gbs}.

\subsection{Is there a quantum connection?}

Despite the fact that 
the laws of black-hole thermodynamics are by now well established, 
many issues remain unclear. It is, for example, not clear to what extent the
black hole entropy 
\begin{equation}
S = { A \over 4}
\label{entrop}\end{equation}
where $A$ is the area of the event horizon,  
can be understood in terms of the statistics 
of a given set of microstates. (We use units such that $c=G=\hbar=1$ throughout the paper.) 
Bekenstein and colleagues have discussed this 
problem in terms of  
a quantised area (see for example \cite{bmuk}). 
Then, in analogy with a typical finite system, a black hole
would have a discrete spectrum. One can argue that
\cite{bmuk}
\begin{equation}
A = n 4 \ln k  	\qquad \mbox{ with } n=1,2,3,... \mbox{ and } k=2,3,4,...
\end{equation}
Comparing this with
\begin{equation}
A = 16 \pi M^2 \rightarrow \Delta A = 32 \pi M \Delta M = 32 \pi M  \omega
\label{corr}\end{equation}
where we have associated the ``energy spacing'' with a frequency 
through $\Delta M = \Delta E = \omega$ (roughly speaking, 
$\omega$ corresponds to a ``transition energy''), one finds that   
the spacing between consecutive states
(for macroscopic black holes, with $M>>\hbar$) will correspond to 
a frequency 
\begin{equation}
 \omega = { \ln k \over 8 \pi M}  
\end{equation}
The standard argument, which favours $k=2$,  
would make the entropy spacing between 
energy levels exactly one ``bit'', which is  attractive from the 
information theory point of view. A key result is that, in this picture 
any radiation will be emitted in multiples of the fundamental 
frequency $\omega$.  Hence, essentially no radiation should be radiated 
with frequencies 
below $\omega$. If true, this is a highly significant conclusion since it suggests that 
the quantum nature of black holes might be observable at the macroscopic level.

As has been demonstrated by Ashtekar and his collaborators \cite{ash1}, one can arrive at 
the same conclusions within the framework of loop quantum gravity (see \cite{ash2} for
a nice introduction). 
This approach is based on the notion of quantum geometry, which means that 
it is natural to ask what the quantum of area might be.
Again, it is possible to draw conclusions from considerations of macroscopic black holes. 
For large black holes it has been shown that \cite{ash1}
\begin{equation}
A = n \gamma_{BI} 4 \pi \sqrt{3}  
\end{equation}
The parameter $\gamma_{BI}$ is an unknown ``natural constant'' called the 
Barbero-Immirzi parameter. It plays an important role because it fixes an ambiguity 
in the theory. If this parameter could be determined by an independent ``experiment''
the theory would become predictive. 
In fact, the Barbero-Immirzi parameter can be fixed by carrying out a calculation 
of the black-hole entropy and comparing to the standard result, Eq.~(\ref{entrop}).
By attributing entropy to microstates (actually nodes of spin 
networks, see \cite{ash2}) it can be shown that 
\begin{equation}
S =  { \ln 2 \over 4 \pi \sqrt{3} \gamma_{BI}} A  
\rightarrow \gamma_{BI} = { \ln 2 \over \pi \sqrt{3}}
\end{equation}
provided that the lowest permissible spin is $1/2$. Again 
the area quantum is $\Delta A = 4 \ln 2$.

So what does this have to do with the black hole QNMs? 
The possible connection between the classical vibrations of a black-hole spacetime
and various quantum aspects has been  discussed for quite some time.
For an early contribution to the debate, see Ref.~\cite{york}.
The recent interest 
in a possible association between the two problems followed a very simple observation. 
A few years ago, Hod \cite{hod} noticed that the numerical results for 
asymptotic Schwarzschild QNMs seemed to suggest that
\begin{equation}
\mbox{Re } \omega M \to  { \ln 3 \over 8\pi} \quad \mbox{ as }  
|\mbox{Im }\omega| \to \infty
\label{res}\end{equation}
If we identify the (real part of the) asymptotic QNM frequency 
with the quantum interstate spacing, we can 
use this value in (\ref{corr}) to get
\begin{equation}
\omega = {\ln 3 \over 8 \pi M} \rightarrow \Delta A = 4 \ln 3
\end{equation}
This is a tantalizing result. It would fit nicely into Bekenstein's 
thermodynamical picture provided that  the 
``fundamental'' frequency is associated with $k=3$ rather than 2.
Furthermore, Dreyer \cite{dreyer} has shown that this would be the prediction of  
loop quantum gravity if it were based on SO(3) rather than SU(2).
In this case the predicted Barbero-Immirzi parameter would be  
\begin{equation}
\gamma_{BI} = { \ln 3 \over 2 \pi \sqrt{2}} 
\end{equation}
There are, of course, problems associated with this change. 
In particular, SO(3) is not favoured because it would seem not to allow coupling to
fermions. Hence, one would need to either explain why fermions
should be excluded \cite{cor}, or provide an alternative derivation of the 
black hole entropy (perhaps using a different statistics for
the quantised area states \cite{poly}).
Various other relevant issues have been discussed in Refs.~\cite{kun,kaul,birm}.
 
On the other hand, we now have a testable prediction.  
If Hod's argument holds,  one should be able to learn something useful 
from, for example, the asymptotic behaviour
of the Reissner-Nordstr\"om QNM frequencies. However, until very recently
only the 
slowly damped QNMs of Reissner-Nordstr\"om black holes 
had been calculated  \cite{gunt,kokk,pam2,leaver2,aas}.

The present paper is motivated by the recent discussion
\cite{motl,motl2,mvdb,bk,neitzke}. 
Our aim is to use
the complex coordinate WKB method \cite{heading,fandf,bam} in its very simplest form to  
determine  asymptotic QNMs for Schwarzschild and Reissner-Nordstr\"om black 
holes. As we will see, this leads to results that agree with 
those of Motl and Neitzke \cite{motl2}. Furthermore, our analysis fills 
several gaps left by their study and extends the discussion
 of the final result considerably.  

\section{Key principles of the complex coordinate WKB analysis}

The equations governing various classes of non-rotating black-hole perturbations 
can  be written \cite{chandra}
\begin{equation}
{ d^2 \psi \over dr_\ast^2} + [ \omega^2 - V(r) ] \psi =0
\end{equation}
where we have assumed that the perturbations depend of time as $e^{-i\omega t}$.
The tortoise coordinate is defined by
\begin{equation}
{ dr_\ast \over dr} = { r^2 \over \Delta} 
\label{tortdef}\end{equation}
where $\Delta$ depends on the spacetime geometry, and is here given by
\begin{equation}
\Delta = r^2 - 2Mr + q^2
\end{equation}
where $M$ is the mass of the black hole and $q$ is its electric charge.
The two solutions to $\Delta = 0$ determine the location of the 
event horizon $r_+ = M + \sqrt{ M^2-q^2}$ 
and, for charged black holes, the inner Cauchy horizon
$r_- = M - \sqrt{ M^2-q^2}$. The definition (\ref{tortdef}) is such that 
the causally attainable spacetime region outside the black hole,
$r_+ < r <\infty$, is mapped onto $-\infty <r_\ast < \infty$. 
The effective potential $V(r)$ is of short range, which means that 
 $\psi \sim e^{\pm i \omega r_\ast}$ both near the 
horizon and at infinity. With our chosen time-dependence
the solution behaving as $e^{i \omega r_\ast}$ represents an 
 outgoing wave at infinity, while $e^{-i \omega r_\ast}$ is a
ingoing wave near the horizon. 

It is  easy to explain one of the main 
difficulties associated with the QNM problem. Suppose we want to calculate
a damped QNM, i.e. a solution for which Im~$\omega M<0$. Then the solution that 
represents outgoing waves at infinity will grow exponentially as 
$r_\ast$ increases. This means that we will need 
exponential precision in order to filter out the ingoing-wave
contribution and impose the desired
boundary condition. This is not a straightforward computational task.
A similar difficulty arises with the boundary condition at the horizon. 
However, the problem becomes straightforward if we analytically continue 
into the complex coordinate plane. Suppose, for example, that we 
analyse the problem along a line such that $\omega r_\ast$ is purely 
real. Then the two asymptotic solutions would be purely oscillatory 
and it would be easy to impose the boundary conditions.
Of course, before we can benefit from this idea we must understand how 
the solutions change under analytic continuation. Fortunately, 
the relevant principles are well known from WKB/phase-integral theory. 
In fact, in this paper we will only use results that were well understood
at least 40 years ago \cite{heading,fandf,bam}. 
The advantage of this kind of analysis is that it does not require the use of 
complicated comparison equations to spot special solutions whose  
analytic properties can be exploited (it is an ``atomic'' description rather than 
a ``molecular'' one).

In order to analyse the black hole problem we prefer to work in the 
complex $r$-plane (this is natural since the tortoise coordinate is multi-valued).
Introducing a new dependent variable
\begin{equation}
\Psi = \left( { \Delta \over r^2} \right)^{1/2} \psi
\end{equation}
we readily rewrite the perturbation equation as
\begin{equation}
{ d^2 \Psi \over dr^2 } + R(r) \Psi = 0
\label{sch}\end{equation}
where 
\begin{equation}
R(r) = \left( { r^2 \over \Delta} \right)^2 \left\{\omega^2 - V(r) + { 1 \over 4} 
\left[ { d \over dr} \left( { \Delta \over r^2 } \right) \right]^2
- { 1 \over 2}{\Delta \over r^2}  { d^2 \over dr^2} \left( { \Delta \over r^2 } \right) 
 \right\}
\label{Rdef}\end{equation}
As is well known, the two WKB solutions to an equation 
of form (\ref{sch}) can be written
\begin{eqnarray}
f_{1,2}^{(t)} (r) = Q^{-1/2} (r) \exp \left[ 
\pm i \int_{t}^r Q(r^\prime) dr^\prime \right]
\label{wkbsol}\end{eqnarray}
with $Q^2=R$. However, under some circumstances it is useful to 
use a slightly different function as basis for the approximation. 
As we will discuss in the next section, we will exercise this 
freedom in our analysis of the asymptotic black hole problem. 
Note that, without loss of generality, the lower limit of integration $t$ 
is customarily taken to be 
one of the zeros of $Q$. Throughout this paper we will indicate the relevant
lower limit of integration by a subscript on $f_{1,2}$ as in (\ref{wkbsol}). 

The zeros and poles of the function $Q$ play a central role in 
any complex coordinate analysis of (\ref{sch}). 
From each simple zero of $Q^2$ emanates three so-called ``Stokes lines''. 
Along each of these contours $Q(r) dr$ is purely imaginary, 
which means that one of the two solutions grows exponentially, 
while the second solution decays, as we move away from $t$. 
In other words, one of the solutions is exponentially dominant
on the Stokes line, while the other solution is sub-dominant.
Analogously, one can define three ``anti-Stokes lines'' associated with each 
simple zero
of $Q^2$. On anti-Stokes lines  $Q(r) dr$ is purely real, 
which means that the two solutions are purely oscillatory. 
As we cross an anti-Stokes line, the dominancy of the two 
functions $f_{1,2}$ changes.
The three-fold symmetry associated with each zero of $Q^2$ 
is clear from Figs.~\ref{fig1} and \ref{fig2}.

Stokes lines are vital for WKB analysis, because it
is in the vicinity of these contours that the solution changes 
character. That is, if the solution is appropriately represented
by a certain linear combination of $f_1$ and $f_2$ in some region of the complex 
$r$-plane, the linear combination will change as the solution is 
extended across a Stokes line. The induced change is not 
complicated: The coefficient of the dominant solution remains 
unchanged, while
the coefficient of the solution which is subdominant on 
the relevant Stokes line picks up a contribution proportional
to the coefficient of the dominant solution.
This is known as the ``Stokes phenomenon'' \cite{stokes}.
The contant of proportionality is known as a ``Stokes constant''. 
This change is necessary for the particular representation (\ref{wkbsol}) 
to preserve the monodromy of the global solution. Terms that are exponentially 
small in one sector of the complex plane may be overlooked. However, in 
other sectors they can grow to exponentially 
dominate the solution. By incorporating the
Stokes phenomenon, we have a formally exact procedure which leads to 
a  proper account of all exponentially small terms.

In the particular case of an isolated simple zero $t$ of $Q^2$ the problem is 
straightforward \cite{heading,fandf}. 
Suppose that the solution in the initial region of
the complex plane is given by
\begin{equation}
\Psi = c f_1^{(t)}
\end{equation}
Then, after crossing a Stokes line
emanating from $t$ (and on which $f_1$ is dominant) the solution becomes
\begin{equation}
\Psi =   c f_1^{(t)} \pm i c f_2^{(t)}
\end{equation}
The sign  depends on whether
one crosses the Stokes line in the positive (anti-clockwise) or 
negative (clockwise) direction.
It is crucial to note that this simple result, i.e. that the Stokes constant
is $\pm i$, only holds when the Stokes line 
emanates from the zero that is  used as lower limit for the phase-integral. 
That is, when we want to 
use the above result to construct an approximate solution valid in various 
regions of the complex plane we must often change the reference point 
for the phase-integral. This leads to the need to evaluate integrals 
of the type
\begin{equation}
\gamma_{ij} = \int_{t_i}^{t_j} Q(r) dr
\end{equation}  
where ${t_i}$ and ${t_j}$ are two simple zeros of $Q^2$. 

A final issue that must be mentioned before we proceed concerns branch cuts. 
In general we need to introduce branch cuts from the simple zeros of $Q^2$ 
in order to ensure that the 
phase integrands remain single-valued. However, one can usually place these
cuts in such a way that they do not affect the analysis. In the following derivations
this is the case.  We will choose the phase of the square-root of $Q^2$ such that 
\begin{equation}
Q = R^{1/2} \sim \omega \quad \mbox{ as } r\to \infty 
\end{equation}
This means that the outgoing-wave solution at infinity is proportional to 
$f_1$ while the ingoing-wave solution at the horizon is 
proportional to $f_2$. 

\section{The Schwarzschild problem}

\subsection{Evaluating the phase-integrals}

In the case of Schwarzschild black holes (when $q=0$)   
there are two classes of gravitational perturbations, usually refered
to as axial and polar \cite{chandra}. We will only consider the axial case here. 
This is, however, no restriction since it has been shown that the 
two cases are isospectral \cite{chandra}. That is, the QNMs are the same 
in both cases. Once we have 
derived the relevant WKB condition for axial gravitational perturbations
we will discuss the case of electromagnetic waves in the Schwarzschild background.  

In the case of axial gravitational perturbations, the effective potential 
is
\begin{equation}
V = { \Delta \over r^2} \left[ {(l(l+1) \over r^2} - { 6 M \over r^3} \right]
\qquad l=2,3,4,...
\end{equation}
From this and Eq.~(\ref{Rdef}) we deduce that 
the function $R$ has two second order poles and four zeros.
Since the zeros are closely associated with the Stokes phenomenon, we need
to know their location, as well as the nature of the Stokes and 
anti-Stokes lines. It is easy to show that,  
when Im~$\omega\to -\infty$ the zeros all approach the origin
of the complex $r$-plane. This allows us to simplify the analysis 
considerably. Expanding 
in a power series near $r=0$ we have
\begin{equation}
R \approx  { r^2 \over 4M^2} 
\left[ \omega^2 - { 15 M^2 \over r^4}\right] \approx  - { 15 \over 4 r^2}
\end{equation}
sufficiently near to the origin. 
Note that this approximation contains no reference
to $l$. As $|\omega| \to \infty$ the $l$-dependent terms are
only higher order corrections. 

Given this behaviour it is easy to 
show that the exact solutions to (\ref{sch}) should behave like
\begin{equation}
\Psi \sim r^{ 1/2\pm 2} \quad \mbox{ as } r\to 0
\end{equation}
Meanwhile, if we take $Q^2=R$ we get
\begin{eqnarray}
Q^{-1/2} &\sim&  r^{-1/2} \\
\int Q dr &\sim& \pm i \sqrt{ 15 \over 4} \ln r 
\end{eqnarray}
which means that
\begin{equation}
f_{1,2} \sim r^{1/2\pm \sqrt{15/4}} 
\end{equation}
In other words, the approximate solutions do not have the correct
behaviour in the vicinity of the origin. However, by choosing
\begin{equation}
Q^2 = R - { 1 \over 4r^2} 
\label{basfun}\end{equation}
we obtain approximate solutions with the desired behaviour near the origin.
This is analogous to the  ``Langer modification''
$l(l+1)\to(l+1/2)^2$ that is used in 
the WKB analysis of radial quantum problems. In principle, one can also adjust 
the approximation near the poles at $r_\pm$ in the black hole 
problem, see for example \cite{pimrev}, but since we are assuming that 
$|\omega|$ is large such alterations would not affect our final result. 
Hence, we will use
\begin{equation}
Q^2 \approx { r^2 \over 4M^2} 
\left[ \omega^2 - { 16 M^2 \over r^4}\right]
\label{q2s}\end{equation} 

In Fig.~\ref{fig1} we show the anti-Stokes and Stokes line geometry
pertaining to (\ref{basfun}) for large $|\omega|$.  
The zeros have been labelled in the same way as in Refs.~\cite{al,gbs}.
In our analysis of the  
QNM problem we will need the phase-integrals
$\gamma_{13}$ and $\gamma_{23}$. 
That is, we need
to evaluate
\begin{equation}
I= \int Q dr \approx \pm \int { r \over 2M} 
\left[ \omega^2 - { 16 M^2 \over r^4}\right]^{1/2} dr 
\end{equation}
where the limits are two neighbouring zeros of $Q^2$. 
Letting $y=\omega r^2/4M$ the zeros map to $-1$ or $1$ and 
this integral becomes
\begin{equation}
I = \pm \int_{-1}^1 \left[ 1 - { 1 \over y^2} \right] dr = 
 \pm \int_{i\pi}^0 { \sinh^2 x \over \cosh x} dx 
= \pm \left[ \sinh x - 2 \arctan e^x \right]_{i\pi}^0 = \mp \pi 
\label{intcal}\end{equation}
Hence we see that, up to the sign, the integrals we need are identical. 
Furthermore, it is easy to show that with our chosen phase for 
$Q$ we obtain
\begin{equation}
\gamma = - \gamma_{13} = - \gamma_{23} = \pi  
\end{equation}
 
\subsection{A WKB condition for asymptotic Schwarzschild QNMs}

We now combine the monodromy argument 
of Motl and Neitzke \cite{motl2} with the standard complex coordinate WKB 
results described in the previous section. 
This will provide a clear argument in support
of the known result for  asymptotic Schwarzchild
black hole QNM frequencies. 
 
\begin{figure}[h] 
\centerline{\epsfysize=7.5cm \epsfbox{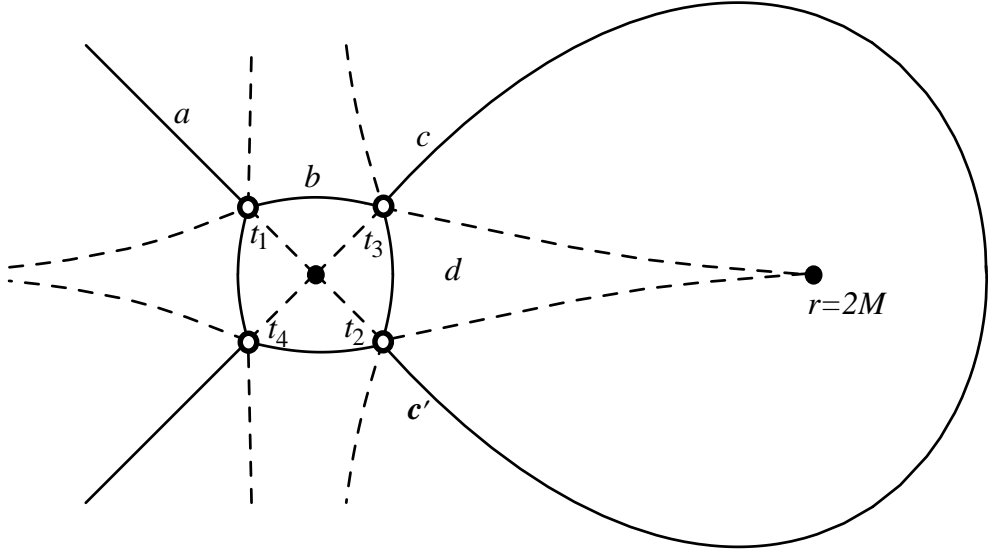}}
\caption{A schematic illustration of the 
Stokes (dashed) and anti-Stokes (solid) lines  for the Schwarzschild problem
in the complex $r$-plane. We have assumed that the
frequency is large and purely imaginary. For frequencies with a small real part, 
the pattern of Stokes and anti-Stokes lines changes only slightly (cf. various figures in 
\cite{al,gbs}). The open circles represent the four zeros of $Q^2$, while the 
filled circles are the two poles (at the origin and the event horizon, respectively).
The outgoing-wave boundary condition at infinity 
is imposed on the anti-Stokes line labelled $a$.
For $\omega = \alpha-i\beta$, with $[\alpha,\beta]>0$ this anti-Stokes line asymptotes
to a straight line at an angle $\arctan \beta/\alpha$, i.e. in the first quadrant of the 
complex $r$-plane.
} 
\label{fig1} 
\end{figure}

For frequencies such that $|\mbox{Im }\omega|>>|\mbox{Re }\omega|$ 
the pattern of Stokes and anti-Stokes lines for the Schwarzschild problem 
is as sketched in Figure~\ref{fig1}. 
Assuming that Re~$\omega M >0$ the outgoing wave boundary condition at spatial infinity
can be analytically continued to the anti-Stokes line labelled $a$ in the figure.  
This issue is discussed in detail in, for example, \cite{pimrev}.
In order to obtain a quantisation condition for highly damped QNMs,
we analytically continue the solution along a closed path 
encircling the pole at the event horizon.  This contour starts out at
$a$, proceeds  along anti-Stokes lines and account for the Stokes phenomenon
associated with the zeros $t_1$, $t_2$ and $t_3$, and eventually ends up at $a$. 
In the analysis we will assume that all zeros and poles of
$Q^2$ are  isolated and can be accounted for individually.

With the chosen phase of
$Q$ the outgoing-wave
solution at point $a$ is
\begin{equation}
\psi_a =  f_1^{(t_1)}
\label{psiA}\end{equation}
Since no Stokes lines cross this contour, 
this solution will not change in character along the anti-Stokes line
that connects point $a$ with the zero $t_1$. 
This means that we can readily
extend the 
solution to the vicinity of $t_1$.
However, if we want to extend the solution to point $b$
on a neighbouring anti-Stokes line 
we must account for the Stokes phenomenon. 
With our choice
of phase for $Q$, the function $f_1^{(t_1)}$ is dominant on the Stokes line
which we must cross in going from $a$ to $b$. 
This means that we will get
\begin{equation}
\psi_b =  f_1^{(t_1)} - i  f_2^{(t_1)}
\end{equation} 
since the Stokes constant for a single well separated zero is $-i$
if we move around the zero in the clockwise direction. 
Changing the lower limit of integration to $t_3$ we get 
\begin{equation}
\psi_b = e^{i\gamma_{13}} f_1^{(t_3)} - i
e^{-i\gamma_{13}}  f_2^{(t_3)} =  
-  f_1^{(t_3)} + i  f_2^{(t_3)}
\end{equation} 
Now extending this solution to point $c$ we do not cross any anti-Stokes lines
so $f_1$ remains dominant. Hence, at $c$ we obtain
\begin{equation}
\psi_c = -  f_1^{(t_3)} + 2i   f_2^{(t_3)}
\label{psiC}\end{equation}
Since we do not cross any Stokes lines in moving from 
$c$ to $c^\prime$, the linear combination (\ref{psiC}) remains a
valid solution. However, we need to note that 
the phase-integral is now evaluated along a contour that loops around the 
pole at $r=2M$. If we replace this integration 
contour by one that lies to the 
left of the pole, we get
\begin{equation}
\psi_{c^\prime} = - e^{i\Gamma}f_1^{(t_3)} + 2i
e^{-i\Gamma} f_2^{(t_3)}
\end{equation}  
where $\Gamma$ is the integral of $Q$ along a contour encircling the pole at $r=2M$
clockwise.

We now want to connect the solution to the point $d$. 
In order to do this we must first ensure that the 
lower limit of the phase-integrals is $t_2$. That is, we use
\begin{equation}
\psi_{c^\prime} = - e^{i(\Gamma+\gamma_{32})}f_1^{(t_2)} + 2i
e^{-i(\Gamma-\gamma_{32})} f_2^{(t_2)}
= e^{i\Gamma}f_1^{(t_2)} - 2i e^{-i\Gamma} f_2^{(t_2)}
\end{equation}
When we step inside the large anti-Stokes lobe in Fig.~\ref{fig1} the dominance of 
$f_1$ and $f_2$ is interchanged. Then, crossing the first of the two 
Stokes lines inside this lobe we obtain
\begin{eqnarray}
\psi_{d} &=& \left[ e^{i\Gamma} + 2e^{-i\Gamma} \right]
f_1^{(t_2)} - 2i e^{-i\Gamma} f_2^{(t_2)}
 = \left[ e^{i\Gamma} + 2e^{-i\Gamma} \right]
e^{i\gamma{23}} f_1^{(t_3)} - 2i e^{-i\Gamma} e^{i\gamma{-23}} f_2^{(t_3)}
\nonumber \\
&=& - \left[ e^{i\Gamma} + 2e^{-i\Gamma} \right] 
f_1^{(t_3)} + 2i e^{-i\Gamma} f_2^{(t_3)}
\end{eqnarray}

Connecting this solution back to $c$, again crossing a Stokes
line where $f_1$ is subdominant, we get (using a bar on $c$ to denote the 
fact that we have encircled the pole at the origin)
\begin{equation}
\psi_{\bar{c}} =  - \left[ e^{i\Gamma} + 4e^{-i\Gamma} \right] 
f_1^{(t_3)} + 2i e^{-i\Gamma} f_2^{(t_3)}
\end{equation}

Now reversing our steps and connecting this solution to $b$ 
we get
\begin{eqnarray}
\psi_{\bar{b}} &=&  - \left[ e^{i\Gamma} + 4e^{-i\Gamma} \right] 
f_1^{(t_3)} -i \left[ e^{i\Gamma} + 2e^{-i\Gamma} \right]  
e^{-i\Gamma} f_2^{(t_3)} \nonumber \\
&=& - \left[ e^{i\Gamma} + 4e^{-i\Gamma} \right] 
e^{i\gamma_{31}} f_1^{(t_1)} -i \left[ e^{i\Gamma} + 2e^{-i\Gamma} \right]  
e^{-i\gamma_{31}} f_2^{(t_1)} \nonumber \\
&=&  \left[ e^{i\Gamma} + 4e^{-i\Gamma} \right] f_1^{(t_1)} 
+i \left[ e^{i\Gamma} + 2e^{-i\Gamma} \right]   f_2^{(t_1)}
\end{eqnarray} 
and (finally) returning to $a$ we have
\begin{eqnarray}
\psi_{\bar{a}} &=& \left[ e^{i\Gamma} + 4e^{-i\Gamma} \right] f_1^{(t_1)} 
+ 2i \left[ e^{i\Gamma} + 3e^{-i\Gamma} \right]   f_2^{(t_1)}
\end{eqnarray} 
Comparing this solution to (\ref{psiA}) we clearly must require that
\begin{equation}
 e^{i\Gamma} + 3e^{-i\Gamma} = 0
\label{schcond}
\end{equation}
in order for the two solutions to be the same. This  would lead to 
\begin{equation}
\psi_{\bar{a}} = e^{-i\Gamma} f_1^{(t_1)}
\end{equation}
from which we see that the clockwise monodromy of this solution
is $e^{-i\Gamma}$.

Let us now perform the same analysis in the vicinity of  the pole at $r=2M$.
With our choice of phase for $Q$, the solution that represents
``ingoing waves'' near the  event horizon is
\begin{equation}
\psi_H = \beta f_2^{(t_3)}
\label{psiH}\end{equation}
For this solution it is trivial to show that 
the clockwise monodromy is (again) $e^{-i\Gamma}$.

Since a necessary condition for 
the two solutions (\ref{psiA}) and (\ref{psiH}) to represent the same global 
solution to the problem is that they have the same monodromy,
we conclude that (\ref{schcond}) is the appropriate 
WKB condition for highly damped 
QNM solutions. 

Finally, using the fact that 
\begin{equation}
\Gamma = \oint Q dr = - 2\pi i \mathop{Res}_{r=2M} Q = - 4\pi i \omega M
\end{equation} 
our WKB condition can be written
\begin{equation}
e^{2i\Gamma} = e^{8\pi\omega M} = -3
\label{wkbcond}\end{equation}
and it immediately follows that 
\begin{equation}
\omega M = { 1 \over 8 \pi} \ln 3 - { i \over 4} \left( n + 
{ 1 \over 2} \right)
\quad \mbox{ as } n \to \infty
\label{schresult}\end{equation}
This is the desired final answer, in complete agreement with 
\cite{motl,motl2,mvdb}. Of course, it is worth noting that our derivation 
is conceptually very simple as it only appeals to  basic WKB
principles.

It is interesting to discuss other classes of black hole perturbations.
In order to do this we note that, had we not used the particular 
value $\gamma=\pi$  in our derivation, we would have arrived at the 
condition
\begin{equation}
e^{8\pi \omega M} = - 1 - 2 \cos  2 \gamma 
\label{gencon}\end{equation}
We can easily  use this condition to determine the asymptotic QNMs for
electromagnetic waves.  

In the case of electromagnetic waves propagating 
in the Schwarzschild geometry the relevant effective potential is
\begin{equation}
V = l(l+1)  { \Delta \over r^4} 
\end{equation}
From this we find, using (\ref{Rdef}) and (\ref{basfun}), that 
\begin{equation}
Q^2 \approx { r^2 \over 4M^2} 
\left[ \omega^2 - { 4 M^2 \over r^4}\right]
\label{Q2E}\end{equation} 
The topology of the problem is still 
represented by Figure~\ref{fig1}, 
and given (\ref{intcal}) we find that 
\begin{equation}
\int Q dr \approx \pm \int { r \over 2M} 
\left[ \omega^2 - { 4 M^2 \over r^4}\right]^{1/2} dr = \pm { \pi \over 2}  
\end{equation}
Hence, we have  $\gamma =\pi/2$ and it follows immediately from 
(\ref{gencon}) that the real part of the QNM frequencies vanishes asymptotically. 

\section{The Reissner-Nordstr\"om problem}

Having verified the known result for Schwarzschild black holes we will
now consider the QNMs of the Reissner-Nordstr\"om
geometry. This problem has recently been discussed by several 
authors, see \cite{motl2,bk,neitzke}, but the implications of the results
still seem far from clear.
The Reissner-Nordstr\"om problem is more complicated than the Schwarzschild
one because of the presence of the inner horizon as well as two
additional zeros of the relevant $Q^2$. Nevertheless, the analysis proceeds
almost exactly as in the previous section. 

In the case of charged black holes, one still only needs to 
consider axial perturbations.  Just as in the Schwarzschild case 
the axial and polar perturbations are isospectral \cite{chandra}.
For
axial perturbations one has two distinct effective potentials, cf. \cite{aas},
\begin{equation}
V_i = { \Delta \over r^2 } \left[ { l(l+1) \over r^2} - { \beta_i \over r^3} 
+ { 4q^2 \over r^4} \right]
\end{equation}
where 
\begin{equation}
\beta_{1,2} =  3M \mp [ M^2 + 4 (l-1)(l+2) q^2] ^{1/2} 
\end{equation}
In the Schwarzschild limit ($q\to 0$) these potentials approach pure 
electromagnetic ($V_1$) and gravitational ($V_2$) perturbations, 
respectively. In the general charged case, the solution corresponds to 
coupled electromagnetic and gravitational waves.  That this should be the case is 
natural
since oscillations of a charged gravitational field will inevitably generate 
electromagnetic waves, and vice versa. Note that the $q\to 0$ limit is, in fact, singular
due to the coalescences of poles and zeros that change the character of the 
function $Q^2$, cf. Figs.~\ref{fig1} and \ref{fig2}.   

\subsection{Evaluating the phase-integrals}

When $q\neq 0$ 
the function $R$ has three second order poles and six zeros
regardless of whether we consider $V_1$ or $V_2$. Furthermore,   
just as in the Schwarzschild case,  the zeros all approach the origin
of the complex $r$-plane when $|\omega|\to \infty$. Again expanding 
in a power series we find
\begin{equation}
R \approx  { r^4 \over q^4} 
\left[ \omega^2 - { 6q^4 \over r^6}\right] \approx  - { 6 \over r^2}
\end{equation}
in the immediate neighbourhood of $r=0$. 
The leading order behaviour is the same for both classes of perturbations ($V_1$ and $V_2$).
Hence, our analysis will hold for both sets of perturbations and the final result 
will be identical in the two cases. 

Repeating the argument from the previous 
section one can show that the choice (\ref{basfun}) still leads to
the WKB solutions having the behaviour expected of the exact solutions
near $r=0$. Thus we will use
\begin{equation}
Q^2 \approx { r^4 \over q^4} 
\left[ \omega^2 - { 25 q^4 \over 4 r^6}\right]
\end{equation} 
As in the previous section we will need the phase-integral
connecting neighbouring zeros of $Q^2$. That is,  we require
\begin{equation}
I= \int Q dr \approx \pm \int { r^2 \over q^2} 
\left[ \omega^2 - { 25 q^4 \over 4 r^6}\right]^{1/2} dr 
\end{equation}
Letting $y=2\omega r^3/5q^2$, and using 
two neighbouring zeros of $Q^2$ as limits, 
the integral becomes  identical to that of the Schwarzschild problem 
(apart from a multiplicative factor), and we find
\begin{equation}
I = \pm { 5 \pi \over 6}  
\end{equation}
Yet again all the integrals we need are the same (up to sign). 
Chosing the phase of $Q$ as in the previous case, and labelling the zeros 
as in Fig.~\ref{fig2}, we have
\begin{equation}
\gamma = - \gamma_{12} = - \gamma_{32} = \gamma_{43} = -\gamma_{54} =  { 5 \pi \over 6}  \ . 
\end{equation}

Let us denote the integral along a contour that encircles the pole at $r_+$ 
(in the negative direction) by $\Gamma_e$. Then
\begin{equation}
\Gamma_e =  \tilde{\gamma}_{25} + \gamma_{54} + \tilde{\gamma}_{43} + \gamma_{32} 
= \tilde{\gamma}_{25} + \tilde{\gamma}_{43} - 2\gamma
\end{equation}
where a tilde indicates that the integral is taken along an anti-Stokes lobe to the right of
either the pole at the event horizon ($\tilde{\gamma}_{25}$) or the pole at the inner
Cauchy horizon ($\tilde{\gamma}_{43}$), cf. Fig~\ref{fig2}.
Similarly, we define
\begin{equation}
\Gamma_i  = - \tilde{\gamma}_{43} + \gamma_{43} =  - \tilde{\gamma}_{43} + \gamma
\end{equation}
These definitions allow us to write
\begin{equation}
\tilde{\gamma}_{43} = - \Gamma_i +  \gamma
\label{til1}\end{equation}
and
\begin{equation}
\tilde{\gamma}_{25} = \Gamma_e + \Gamma_i + \gamma  
\label{til2}\end{equation}

The two integrals $\Gamma_e$ and $\Gamma_i$ are readily evaluated
using the residue theorem. We find
\begin{equation}
\Gamma_e = -2\pi i \mathop{Res}_{r=r_+} Q = - \pi i { \omega r_+^2 \over 
\sqrt{ M^2 - q^2}   } =  - \pi i { \omega M \over \kappa} ( 1+\kappa)^2
\label{Ge}\end{equation}
where we have defined the dimensionless parameter
\begin{equation}
\kappa = \sqrt{1 - {q^2 \over M^2}  } 
\end{equation}
The integral around the inner horizon is
\begin{equation}
\Gamma_i = -2\pi i \mathop{Res}_{r=r_-} Q =  \pi i { \omega r_-^2 \over 
\sqrt{ M^2 - q^2}   } =   \pi i { \omega M \over \kappa} ( 1-\kappa)^2
\label{Gi}\end{equation}

\subsection{The WKB condition for Reissner-Nordstr\"om QNMs}

\begin{figure}[h] 
\centerline{\epsfysize=7.5cm \epsfbox{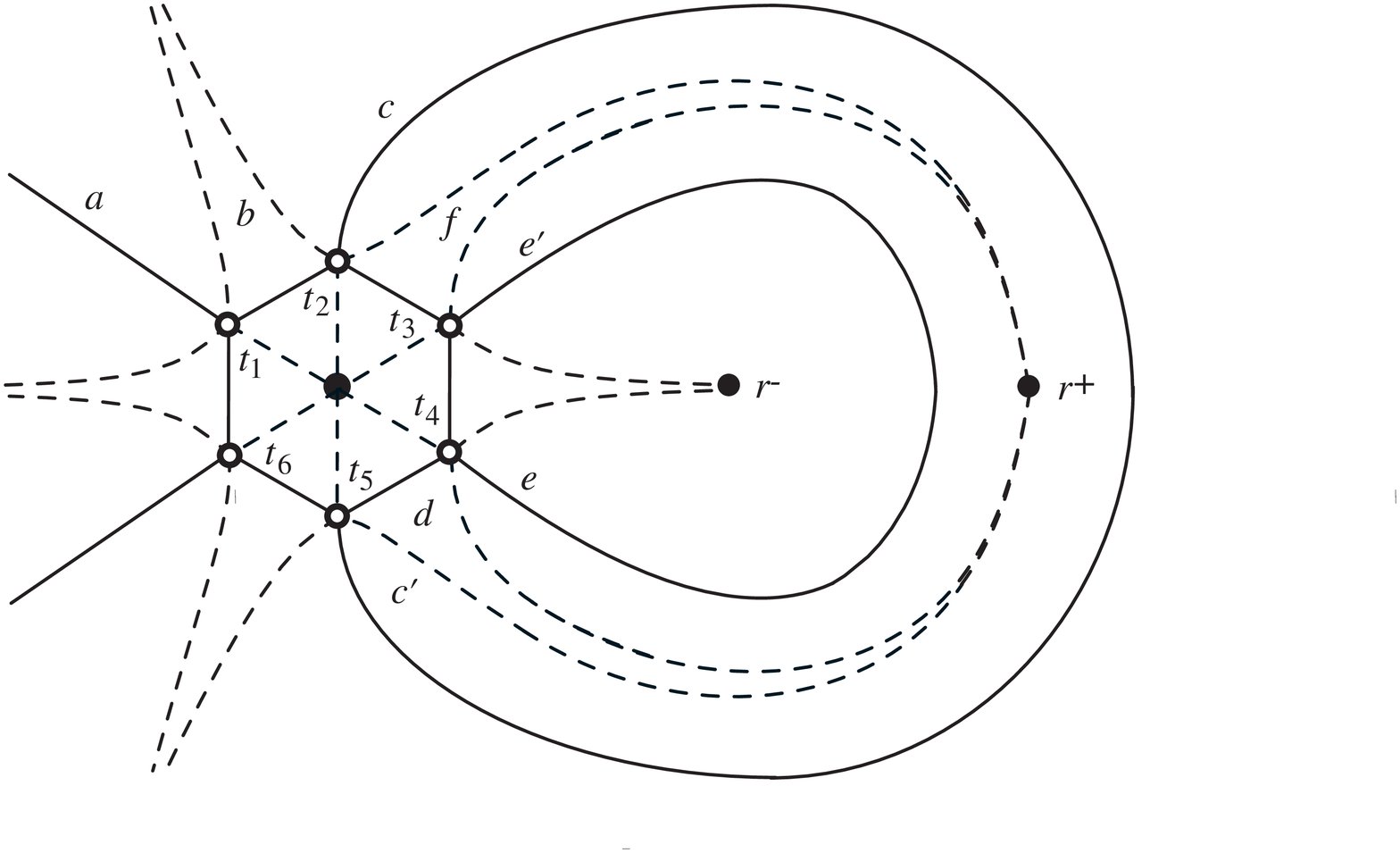}}
\caption{Stokes and anti-Stokes lines for Reissner-Nordstr\"om problem. Just as Fig.~\ref{fig1},
this is a schematic picture based on a purely imaginary frequency. The three poles in the problem
are represented by filled circles and correspond to, from left to right, the origin, the 
inner horizon and the event horizon. The six zeros of $Q^2$ are shown as open circles. 
} 
\label{fig2} 
\end{figure} 

The pattern of Stokes and anti-Stokes lines for the Reissner-Nordstr\"om problem 
for $|\mbox{Im }\omega|>>|\mbox{Re }\omega|$ 
is shown in Figure~\ref{fig2}. 
Introducing the outgoing-wave boundary condition on the appropriate anti-Stokes line
(at point $a$ in Fig.~\ref{fig2}), 
just as in the Schwarzschild problem, and then 
moving around the pole at $r=r_+$, taking full account of all involved
Stokes phenomena (assuming the zeros are well isolated and can be treated
individually),  we arrive at the following WKB condition for highly damped QNMs
\begin{equation}
  e^{2i\Gamma_e}  = 1 -   \left( 1 +  e^{-2i\gamma} \right) 
\left( 1 +  e^{2i\gamma} \right) \left( 1 + 
 e^{-2i\Gamma_i} \right)
\end{equation}
(the complete derivation is provided in the Appendix). 
Since we know that $\gamma=5\pi/6$ this becomes
\begin{equation}
 e^{2i\Gamma_e}  = 1 - 2 \left(1 + \cos { 5 \pi \over 3}  \right)  \left( 1 + 
 e^{-2i\Gamma_i} \right) = -2 - 3  e^{-2i\Gamma_i}
\label{RNcon}\end{equation}
which can be shown to be 
identical to the condition  recently 
derived by Motl and Neitzke \cite{motl2} using matched asymptotics.
That this condition agrees well with the numerical solution to the QNM problem, 
for various given overtones, has been shown
by Berti and Kokkotas \cite{bk}.

\subsection{Approaching the Schwarzschild limit} 

It is interesting to consider what happens
when the black hole approaches the Schwarzschild limit, 
 since (\ref{RNcon}) is clearly
at variance with the result (\ref{wkbcond}) for Schwarzschild black holes. 
For $|\omega| \to \infty$ then $q\to 0$, we have
\begin{equation}
\Gamma_e \approx - 4\pi i \omega M \left(1 + { q^4 \over 16M^4}  \right)  
\end{equation}
and
\begin{equation}
\Gamma_i \approx \pi i \omega M { q^4 \over 4M^3}   
\end{equation}
Thus (\ref{RNcon}) predicts that we ought to have
\begin{equation}
e^{8\pi \omega M} = - 5 
\end{equation}
Yet the Schwarzschild calculation predicted that the right-hand side should 
be $-3$!
The reason for the discrepancy is, however, easy to explain. As we have already mentioned
the Schwarzschild limit is singular. At our level of approximation 
one cannot move from the topology
of Fig.~\ref{fig2} to that of  Fig.~\ref{fig1} in a non-trivial way. This would 
require a uniform approximation involving the coalescence of two zeros and two poles.

The asymptotic behaviour of the QNMs of a charged black hole is always 
given by (\ref{RNcon}), and hence corresponds to the real part of the 
QNM frequencies approaching $\ln 5 /8\pi$ in the limit of infinite damping. 
But there is also likely to be an intermediate 
regime in which the highly damped QNMs more resemble the Schwarzschild 
result, i.e. where Re~$\omega M \approx \ln3/8\pi$.
As the Schwarzschild limit is approached, this latter regime tends to dominate,
with the true Reissner-Nordstr\"om asymptotic behaviour being relevant only
for extremely rapidly damped QNMs. That this is the case can be understood by the 
following argument: Let us consider a black hole with an infinitesimal charge and 
QNMs such that $|\mbox{Im }\omega M|>> |\mbox{Re }\omega M|$. The main difference 
between this problem and the Schwarzschild case is the presence of the double pole 
associated with the inner horizon $r_-$, and two additional zeroes of the function $R$,
defined by (\ref{Rdef}).  As the imaginary part of the QNMs increases, all six zeros of 
$R$ move towards the double pole at the origin. Our analysis of the problem is only relevant 
when the topology is that illustrated in Fig.~\ref{fig2}, i.e. when the pole at $r_-$ lies outside 
the circle on which the six zeroes are located. For infinitesimally charged black holes 
there will also exist a regime where the topology of the 
the four zeroes  already present  in the Schwarzschild case are essentially
unchanged. For this to be true, we need to have
\begin{equation}
|t| \approx \left(M/\omega \right)^{1/2} 
\end{equation}
At the same time the pole at $r_-$, and the two additional zeroes that come into existence 
when the black hole attains charge (and which emerge from the origin together with $r_-$), 
must lie well inside the circle of ``Schwarzschild'' zeroes. This corresponds to
\begin{equation}
|r_-|<<|t| \rightarrow q^2 << 2M \left(M/\omega \right)^{1/2} 
\end{equation}
If these conditions are met one would expect the QNMs to be similar to the
Schwarzschild ones. Figure~\ref{fig3} is a schematic illustration of the two 
regimes for highly damped Reissner-Nordstr\"om QNMs.  
Even though it is easy to explain the behaviour in principle,  
it is not straightforward to extend our analysis
into a consistent scheme for calculating the QNMs in the intermediate regime. 
The main reason for this is the need to evaluate the phase-integral $\gamma$. 
In the intermediate regime we can no longer make use of the power series expansion 
around the origin
that led to (\ref{q2s}). Such an expansion is only valid up to the nearest 
pole, i.e. $r_-$. Resolving this issue, and determining the QNMs also in this
intermediate regime, may be an interesting problem but we do not consider 
it further  here.  

\subsection{Solving the QNM condition}

In order to discuss the solutions to the Reissner-Nordstr\"om 
QNM condition, it is useful to rewrite (\ref{RNcon}) as
\begin{equation}
e^{8\pi \omega M} = - 3 - 2 e^{-2\pi\omega M (1-\kappa)^2/\kappa}
\end{equation}
where we have used (\ref{Ge}) and (\ref{Gi}).
From this we easily see that in the extremely charged case, as
$\kappa \to 0$, we regain the Schwarzschild result \cite{neitzke}.
That is, we get
\begin{equation}
\mbox{Re } \omega M \to { \ln 3 \over 8 \pi} \quad \mbox{ as  } q\to M
\end{equation}
It is, however, not clear to what extent this result is relevant. 
After all, the extreme Reissner-Nordstr\"om limit is singular in the 
sense that the two poles at $r_\pm$ coalesce as $q\to M$. 
This means that the topology is no longer that illustrated in Fig.~\ref{fig2},
and hence the extreme Reissner-Nordstr\"om case would 
require a separate analysis.  

In order to analyse the general case, we introduce 
\begin{equation}
y=8\pi \omega M \qquad \mbox{ and } \quad k= { (1-\kappa)^2 \over 4 \kappa}
\label{yydef}\end{equation}
The condition can then be written as
\begin{equation}
e^y = -3 -2 e^{-ky} \ .
\end{equation}
In general, this condition must be solved numerically. One way to do this is to 
first separate the real and imaginary parts of the equation. Letting 
$y=\alpha -i\beta$ we obtain the two equations
\begin{eqnarray}
e^\alpha \cos \beta &=& -3 -2 e^{-k\alpha} \cos k\beta \label{rel}\\
e^\alpha \sin \beta &=& 2 e^{-k\alpha} \sin k \beta \label{ima}
\end{eqnarray}
These equations are very useful. First of all we see that, 
if we want the solution to 
be periodic in $\beta$, we must require simultaneously 
\begin{equation}
\beta \to \beta + 2n\pi \quad \mbox{ and }  \quad \quad k(\beta + 2n\pi) = k\beta +  2m \pi
\end{equation}
where $n$ and $m$ are integers. That is, 
periodicity in Im~$\omega$ is only possible if 
\begin{equation}
k = {m \over n}
\end{equation}
Since $0\le k \le \infty$ we see that 
we will pass through all (positive) rational values for $k$
as the charge of the black hole is varied. Hence, there will 
be an infinite set 
of cases where the solution is periodic in the 
imaginary part. However, whenever $k$ is not a rational number,
the spectrum ceases to 
be periodic in the imaginary part. This is a significant observation because it 
illustrates that the asymptotic Reissner-Nordstr\"om spectrum is generally 
very different from that of a  Schwarzschild black hole. 
One might wonder if the inclusion of higher order terms in the approximation 
would reimpose periodicity in Im~$\omega$ for irrational values of $k$. 
However, this seems unlikely since it would require a surprising
``conspiracy'' between these terms. The difficulty (due to truncation error)
of representing exactly rational numbers would pose significant challenges 
for a numerical verification 
of this result. 

\begin{figure}[h] 
\centerline{\epsfysize=10cm \epsfbox{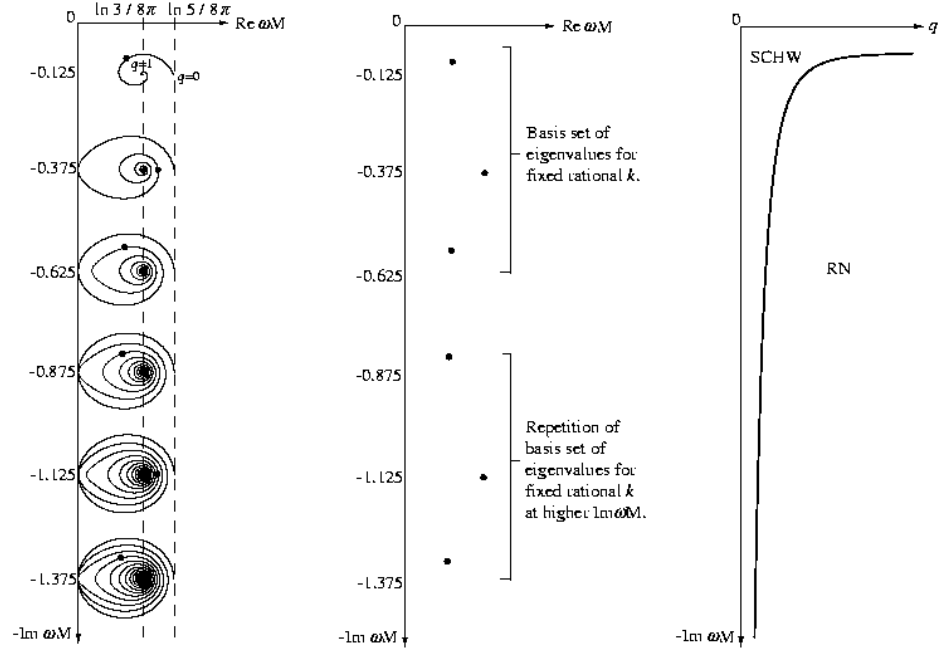}}
\caption{Left panel: A schematic illustration of the behaviour of the first few solutions 
to the Reissner-Nordstr\"om QNM condition (\ref{RNcon}). As the charge is increased the 
QNM frequencies spiral towards $\ln 3/8\pi - (2p+1)i/8$. The figure illustrates 
how the spirals tighten with increasing damping. Generally, the asymptotic QNMs correspond to 
the high damping limit of the figure. Middle panel: 
In the case of fixed rational $k=m/n$, the spectrum is divided into repetitions of basis sets
of eigenvalues from low-lying states (indicated by dots). 
The number of modes in the basis set increases with $n$. For irrational $k$ this 
periodicity is destroyed.    
Right panel: 
A schematic illustration of the two asymptotic regimes of the Reissner-Nordstr\"om problem, 
RN and SCHW in the figure.
In the limit of infinite damping the QNMs are determined by (\ref{RNcon}), but there
also exists an intermediate ``Schwarzschild-like'' regime . The boundary between the two regimes is
(roughly) given by $q^2 \approx 2M \left(M/\omega \right)^{1/2}$. 
} 
\label{fig3} 
\end{figure}

For $m$ and $n$ integers we can introduce $z=e^{y/n}$ to get
\begin{equation}
z^{n+m} + 3z^m + 2 = 0
\end{equation}
which obviously has $n+m$ roots (one of these
polynomial cases was discussed in  \cite{neitzke}).
These roots lead to 
\begin{equation}
8\pi \omega M = y = n \ln |z| + in \arg z + 2n p \pi i \quad p = 0, 1, 2, ...
\label{periods}\end{equation}
Only some of these roots will be compatible
with the derivation of our WKB condition since we assumed that Re~$\omega M\ge0$
at the outset.
The permissible roots correspond to the smallest basis set required to recontruct 
the spectrum. Repetitions of these 
roots on other sheets of $z$ (different $k$), for large $p$ generate the 
asymptotic spectrum, cf. figure~\ref{fig3}. 
Hence, it is interesting to consider some of these polynomial solutions. 
In Table~\ref{tab1} we give the roots that are compatible with the 
derivation of our QNM condition for some selected values of $k=m/n$.

The data given in Table~\ref{tab1} provide the basis needed to generate the asymptotic QNM spectrum
in the simplest periodic cases. As shown in the table, there are $n$ distinct QNMs in each basis 
set. These yield the asymptotic spectrum for large values of $p$, cf. (\ref{periods}).
These results are only valid in the regime where the topology of the problem is that 
shown in Figure~\ref{fig2}. From our discussion of the Schwarzschild limit above, we deduce that 
this is the case when 
\begin{equation}
|\mbox{Im }\omega M |>> 4 (M/q)^4 
\end{equation}

It is  relevant to ask if there are any values of $k$ for which we have purely
imaginary frequencies. One can show that $\alpha=0$ is only compatible
with (\ref{rel}) and (\ref{ima}) provided that
\begin{equation}
\sin \beta = 0 \quad \mbox{ and } \cos \beta = -1
\end{equation}
That is, we must have
\begin{equation}
k = { 2m+1 \over 2n+1}  
\end{equation}
where $n$ and $m$ are integers. One of these roots (for $m=1$ and $n=3$) is present in 
Table~\ref{tab1}. These purely imaginary solution may be the cause of some confusion 
because it is debatable whether they are compatible with the assumptions made in 
the derivation of our QNM condition. In particular, it is not clear to what extent these 
solutions are relevant for a discussion of ``purely outgoing wave'' solutions. 
Furthermore, one has to be careful because only some of the 
purely imaginary roots belong to modes that can be traced back to a Schwarzschild 
QNM. The  condition (\ref{RNcon}) was derived assuming that the real part of 
the frequency was positive. Yet, a numerical solution of the condition also yields
roots with a negative real part. These are not compatible with the underlying assumptions 
and should be discarded as unphysical. These unphysical roots also form spirals as 
$q$ is varied. Some of the purely imaginary roots of the polynomial belong to these
presumed unphysical solutions.  

By complementing the polynomial roots with a numerical solution of 
(\ref{RNcon}), cf. Figure~\ref{fig3},  we can further elucidate the behaviour
of the asymptotic charged black hole QNMs. The numerical solutions  illustrated in 
Figure~\ref{fig3} clearly show the spiral nature of the QNMs. Furthermore, we see that the 
number of times that each spiral touches the imaginary axis increases with the 
``order'' of the mode. It is clear that, as Im~$\omega M \to -\infty$ the spirals tighten, i.e.
the QNM frequency becomes exponentially sensitive to variations in $q$. Whether this is a hint that 
the problem becomes probabilistic in the limit of infinite damping is not clear. 

\begin{table}
\begin{tabular}{|c|c|c|l|l|l|l|}
\hline
$m$ & $n$  	& $q/M$ & \multicolumn{4}{c}{$\omega M$}   \\
\hline
$\infty$ & --- 	&  $1^\dag$	   & \multicolumn{3}{l}{$\ln 3/8\pi - i/8=0.04371-i/8$ } & \\
\hline
8 & 1		& 0.9996 & $0.04372-i/8$	& & & \\
7 & 1		& 0.9994 & $0.04370-i/8$ & 	 & & \\
6 & 1		& 0.9993 & $0.04375-i/8$ &&& \\
5 & 1 		& 0.9990 & $0.04360-i/8$	&  & & \\
4 & 1 		& 0.9984 & $0.04403-i/8$	& & & \\
7 & 2 		& 0.9980 & $0.04374-0.12556i$	& $0.04374-0.37444$ & & \\
3 & 1		& 0.9974 & $0.04263-i/8$	& & & \\
8 & 3 		& 0.9968 & $0.04311-0.12364i$	& $0.04499-3i/8$ & $0.04311-0.62636i$  & \\
5 & 2 		& 0.9965 & $0.04392-0.12334i$	& $0.04392-0.37666i$ & & \\
7 & 3		& 0.9960 & $0.04483-0.12348i$	& $0.04128-3i/8$ & $0.04483-0.62652i$ & \\
2 & 1		& 0.9949 & $0.04623-i/8$	& & & \\ 
7 & 4 		& 0.9936 & $0.04634-0.12707i$	& $0.04129-0.37873i$	& $0.04129-0.61127i$ & $0.04634-0.87293i$ \\
5 & 3 		& 0.9931 & $0.04607-0.12790i$ 	& $0.03788-3i/8$ & $0.04607-0.62290i$  &  \\
3 & 2		& 0.9919 & $0.04484-0.12969i$ & $0.04484-0.37031i$ & & \\
4 & 3		& 0.9903 & $0.04222-0.13140i$ & $0.04860-3i/8$ & $0.04222-0.61860i$ & \\
5 & 4 		& 0.9893 & $0.04008-0.13200i$ 	& $0.04815-0.37830i$ & $0.04815-0.62170i$ & 
$0.04008-0.86800i$ \\
1 & 1 		& 0.9852 & $\ln 2/8\pi-i/8$ 	& & &  \\
3 & 4		& 0.9780 & $0.03849-0.11227i$	& $0.05121-0.36969i$ & $0.05121-0.63031i$ & $0.03849-0.88773i$ \\
2 & 3 		& 0.9743 & $0.04193-0.11162i$	& $0.05331-3i/8$ & $0.04193-0.63838i$ & \\
1 & 2	 	& 0.9634 & $0.04817-0.11199i$ 	& $0.04817-0.38801i$  & & \\
1 & 3  		& 0.9428 & $0.05391-0.11419i$	& $-3i/8$& $0.05391-0.63581i$ & \\ 
1 & 4  		& 0.9242 & $0.05664-0.11598i$	& $0.03493-0.35261i$ 	& $0.03493-0.64739i$ & $0.05664-0.88402i$\\
\hline
--- & $\infty$  & $0^\dag$ &   \multicolumn{3}{l}{$\ln 5/8\pi -i/8=0.06404-i/8 $ } &  \\
\hline\end{tabular}
\caption{A sample of roots in cases where the asymptotic spectrum is periodic in the imaginary part. 
These roots for the basis sets out of which the highly damped QNM spectrum is constructed. 
$^\dag$We include the predictions of (\ref{RNcon}) for both the 
Schwarzschild limit and the extremely charged
black hole limit despite these limits being ``singular'', cf. the discussion in the main text.}
\label{tab1}\end{table}

\section{Discussion}

In this paper we have provided WKB results for highly damped quasinormal modes
of Schwarzschild and Reissner-Nordstr\"om black holes. In the Schwarzschild 
case, we have verified previous results for gravitational perturbations. In particular, 
we have provided a clear derivation of the fact that the real part of the
quasinormal mode frequencies approaches Re~$\omega M = \ln 3/8\pi$ asymptotically, 
cf. Eq.~(\ref{schresult}).
We have also shown that the real part of the asymptotic modes vanishes
for electromagnetic waves.  
For Reissner-Nordstr\"om black holes we have verified the QNM condition 
recently obtained by Motl and Neitzke \cite{motl2}, our Eq.~(\ref{RNcon}). 
In addition, we 
have shown that this result holds to leading order, 
not only for the class of perturbations that limits 
to pure gravitational waves as $q\to0$, but also for the class that limits
to pure electromagnetic waves. That these two classes of perturbations are 
isospectral for highly damped modes was not known previously. 
Perhaps the most relevant result concerns the fact that the 
asymptotic QNM frequencies of charged black holes are
in general not periodic in the imaginary part. This means that the 
asymptotic spectrum is significantly more complicated than in the 
Schwarzschild case. 
We have also explained the breakdown of the analysis in the 
Schwarzschild limit.
The WKB approach that we have used should be applicable to many similar 
problems. In particular, it should be possible to 
use our approach to study the Kerr problem \cite{bk,hod2}. 

Before we conclude the paper, let us return to the question of a 
possible link between the QNMs and quantum gravity.
Do the present results shed further light on this 
association, or does it now seem as if Hod's original suggestion was based on
a misleading coincidence? Unfortunately, the available results do not 
provide a clear answer to this question. Hence, we
conclude this paper with some speculations. 

First, the original result
for gravitational perturbations of Schwarzschild black holes holds. 
If one accepts the proposed correspondence between the QNM frequencies
and a ``transition energy''  $\Delta M$ one finds that 
the quantum of area should be $\Delta A= 4\ln 3$. How do results for other 
perturbing fields fit into this picture? For example, what about the fact that 
the real part of the electromagnetic QNMs vanishes asymptotically? 
Although interesting in its own right, this result is probably not relevant
for the discussion. A natural reason for this would be the fundamental difference 
between gravitational and electromagnetic perturbations of Schwarzschild black holes. 
While electromagnetic perturbations correspond to waves propagating in a fixed 
background, the gravitational waves  represent oscillations of the 
spacetime itself. It would perhaps not be surprising if results for the former problem 
tell us little about the quantum levels of the black hole. Could
a study of massive fields allow us to make progress?

The situation is even more complicated when we turn to the Reissner-Nordstr\"om problem.
How can our results be understood? At first sight it
would be tempting to suggest that the generally nonperiodic nature of the 
asymptotic QNM spectrum, with no unique Re~$\omega M$,  
provides an argument against any quantum correspondence. 
However, this conclusion might be premature. 
One reason for this is that a general perturbation of a 
Reissner-Nordstr\"om black hole corresponds to a mixture of electromagnetic and gravitational waves. 
Would it be too surprising if this mixing were to prevent a simple
correspondence argument?
Moreover, we have the unexpected 
result that we retain the same result for the asymptotic QNMs in both the 
Schwarzschild and the extreme Reissner-Nordstr\"om limits. Is this telling us something 
profound or is it (again) a mere coincidence? 

Suppose we accept the association between the asymptotic QNMs 
and the quantum area $\Delta A$ . What would this imply for 
a general charged black hole? The area of a charged black hole is given by 
\begin{equation}
A = 4 \pi r_+^2  
\end{equation}
From this we readily get (using $\kappa$ as defined in Eq.~(\ref{yydef})) 
\begin{equation}
\Delta A  =  { 8 \pi M(1+\kappa)^2  \over \kappa } \Delta M - 
{ 8\pi q (1+\kappa) \over \kappa} \Delta q = { 8\pi r_+^2 \over M\kappa}\Delta M  
- { 8\pi q r_+ \over M \kappa} \Delta q
\label{dA}\end{equation}
From this relation it is clear that, in general, knowledge of $\Delta M$ alone is not sufficient
to ``predict'' $\Delta A$. We also need to disentangle the 
electromagnetic waves from the gravitational ones. 

In the absence of more information we can play a simple game. Let us assume that 
 $\Delta A$ is indeed universal and thus remains as in the Schwarzschild 
case ($=4\ln 3$). Then we can infer from (\ref{dA}) that
\begin{equation}
q \Delta q = r_+ \Delta M - { \kappa M \ln 3 \over 2 \pi r_+} 
\end{equation} 
If we associate an ``energy'' with a given oscillation frequency
we have $\Delta M = \omega$ and thus
\begin{equation}
q\Delta q = r_+ \omega - { \kappa M \ln 3 \over 2 \pi r_+} 
\label{qdq}\end{equation}
From this relation we can, given any QNM,  compute the value of $\Delta q$
required for us to obtain the same unique $\Delta A$. 
In the particular case of 
extremely charged black holes we would get (since $\kappa\to 0$ as $q\to M$) 
\begin{equation}
\Delta q =  r_+ \omega = \omega M = { \ln 3 \over 8 \pi} = \Delta M   
\end{equation}
We can also use (\ref{qdq}) as a measure of the degree to 
which a perturbation is ``electromagnetic'' or ``gravitational''. 
To what extent this kind of analysis makes sense is, of course, an open question.

\section*{Appendix: The Reissner-Nordstr\"om calculation}

In this Appendix we provide the complete derivation of the WKB condition 
for highly damped Reissner-Nordstr\"om QNMs. 

The outgoing-wave
solution at point $a$ is
\begin{equation}
\psi_a =  f_1^{(t_1)}
\label{RpsiA}\end{equation}
Moving around the pole at $r=r_+$, cf. Fig.~\ref{fig2}, taking full account of all involved
Stokes phenomena (assuming that the zeros are well isolated) we get

\begin{equation}
\psi_b =  f_1^{(t_1)} - i  f_2^{(t_1)} =  e^{-i\gamma} f_1^{(t_2)} - i e^{i\gamma} f_2^{(t_2)}
\end{equation} 

\begin{equation}
\psi_c =  e^{-i\gamma} f_1^{(t_2)} - i \left[ e^{i\gamma}+  e^{-i\gamma} \right] f_2^{(t_2)}
\end{equation} 

\begin{equation}
\psi_{c^\prime} =  
e^{i(\Gamma_e+\Gamma_i)} f_1^{(t_5)} - i \left[ 1 +  e^{-2i\gamma} \right] 
e^{-i(\Gamma_e+\Gamma_i)} f_2^{(t_5)} 
\end{equation} 
where we have used (\ref{til2}), and then 
\begin{eqnarray}
\psi_{d} &=&  
\left\{ e^{i(\Gamma_e+\Gamma_i)} + \left[ 1 +  e^{-2i\gamma} \right] 
e^{-i(\Gamma_e+\Gamma_i)} \right\} f_1^{(t_5)} - i \left[ 1 +  e^{-2i\gamma} \right] 
e^{-i(\Gamma_e+\Gamma_i)} f_2^{(t_5)} \nonumber \\
&=&  
\left\{ e^{i(\Gamma_e+\Gamma_i-\gamma)} + e^{-i\gamma}\left[ 1 +  e^{-2i\gamma} \right]
e^{-i(\Gamma_e+\Gamma_i)} \right\}  f_1^{(t_4)} - i \left[ e^{i\gamma} +  e^{-i\gamma} \right] 
e^{-i(\Gamma_e+\Gamma_i)} f_2^{(t_4)}
\end{eqnarray} 

\begin{eqnarray}
\psi_{e} &=&  
\left\{ e^{i(\Gamma_e+\Gamma_i)} + \left( 1 +  e^{2i\gamma}\right) \left(1 +  e^{-2 i\gamma} \right) 
e^{-i(\Gamma_e+\Gamma_i)} \right\}  e^{-i\gamma} f_1^{(t_4)} \nonumber \\
&-& i \left( 1 +  e^{-2i\gamma} \right)
e^{-i(\Gamma_e+\Gamma_i-\gamma)} f_2^{(t_4)}
\end{eqnarray} 

Moving around the pole at the inner horizon (to $e^\prime$), 
the solution does not change because no
Stokes lines are crossed, but changing the lower limit of integration back to $t_3$ we get
\begin{eqnarray}
\psi_{e^\prime} &=&  
\left\{ e^{i\Gamma_e} +  \left( 1 +  e^{2i\gamma} \right)\left( 1 +  e^{- 2i\gamma} \right)
e^{-i(\Gamma_e+2\Gamma_i)} \right\}  f_1^{(t_3)} - i \left( 1 +  e^{-2i\gamma} \right)
e^{-i\Gamma_e} f_2^{(t_3)}
\end{eqnarray} 

Then
\begin{eqnarray}
\psi_{f} &=&  
\left\{ e^{i\Gamma_e} + \left( 1 +  e^{-2i\gamma} \right) \left[ 1 + 
\left( 1 +  e^{2i\gamma} \right)
 e^{-2i\Gamma_i}  \right] 
e^{-i\Gamma_e}\right\}  f_1^{(t_3)} 
- i \left( 1 + e^{-2i\gamma} \right)
e^{-i\Gamma_e} f_2^{(t_3)}
 \nonumber \\
&=& e^{-i\gamma} 
\left\{ e^{i\Gamma_e} + \left(1 +  e^{-2i\gamma} \right) \left[ 1 + 
\left( 1 +  e^{2i\gamma} \right)
 e^{-2i\Gamma_i}  \right] 
e^{-i\Gamma_e}\right\}  f_1^{(t_2)} 
\nonumber \\
&-& i \left[ 1 +  e^{-2i\gamma} \right] 
e^{-i(\Gamma_e-\gamma)} f_2^{(t_2)}
\end{eqnarray} 

\begin{eqnarray}
\psi_{\bar{c}} &=&
e^{-i\gamma} \left\{ e^{i\Gamma_e} + 
\left( 1 +  e^{- 2i\gamma} \right) \left[ 1 + e^{2i\gamma} + 
\left( 1 +  e^{2i\gamma} \right)
 e^{-2i\Gamma_i} \right] 
e^{-i\Gamma_e} \right\}  f_1^{(t_2)} \nonumber \\ 
&-& i \left( 1 +  e^{-2i\gamma} \right) 
e^{-i(\Gamma_e-\gamma)} f_2^{(t_2)}
\end{eqnarray} 

\begin{eqnarray}
\psi_{\bar{b}} &=&
e^{-i\gamma} \left\{ e^{i\Gamma_e} + 
\left( 1 +  e^{- 2i\gamma} \right) \left[ 1 + e^{2i\gamma} + 
\left( 1 +  e^{2i\gamma} \right)
 e^{-2i\Gamma_i} \right] 
e^{-i\Gamma_e} \right\}
f_1^{(t_2)} \nonumber \\ 
&+& i e^{-i\gamma} \left\{ e^{i\Gamma_e} + 
\left( 1 +  e^{- 2i\gamma} \right) \left[ 1  + 
\left( 1 +  e^{2i\gamma} \right)
 e^{-2i\Gamma_i} \right] 
e^{-i\Gamma_e} \right\}  f_2^{(t_2)}\nonumber \\
 &=& A  f_1^{(t_1)} + i B f_2^{(t_1)}
\end{eqnarray}
where
\begin{equation}
A =  e^{i\Gamma_e} + \left( 1 +  e^{-2i\gamma} \right) 
\left( 1 +  e^{2i\gamma} \right) \left( 1 + 
 e^{-2i\Gamma_i} \right)
e^{-i\Gamma_e}
\end{equation}

\begin{eqnarray}
B &=&  e^{-2i\gamma} \left\{ 
e^{i\Gamma_e} + \left( 1 +  e^{-2i\gamma} \right)\left[ 1 + 
\left( 1 +  e^{2i\gamma} \right)
 e^{-2i\Gamma_i}  \right] 
e^{-i\Gamma_e} \right\} \nonumber \\
&=& A e^{-2i\gamma}
-  \left( 1 + e^{-2i\gamma} \right) e^{-i\Gamma_e} 
\end{eqnarray}

Finally returning to the starting point $a$, we have 
\begin{equation}
\psi_{\bar{a}} = A f_1^{(t_1)} + i (A + B) f_2^{(t_1)}
\end{equation}

Thus we see that, in order for the coefficient of $f_2$ to vanish we must 
require
\begin{eqnarray}
A + B = A + A e^{-2i\gamma}
-  \left( 1 + e^{-2i\gamma} \right) e^{-i\Gamma_e} = 0 
\end{eqnarray} 
That is, we should have
\begin{equation}
A = e^{-i\Gamma_e} 
\end{equation}
which is not too surprising given the result in the Schwarzschild case. 
As in that problem, one readily shows that the (clockwise) monodromy of the solution that 
is purely ingoing at the event horizon is also $ e^{-i\Gamma_e}$. Hence, our calculation is 
consistent. Given the definition of $A$ the WKB condition for highly 
damped Reissner-Nordstr\"om QNMs becomes  
\begin{equation}
  e^{2i\Gamma_e}  = 1 -   \left( 1 +  e^{-2i\gamma} \right) 
\left( 1 +  e^{2i\gamma} \right) \left( 1 + 
 e^{-2i\Gamma_i} \right)
\end{equation}

\end{document}